\documentclass[english,aps,pra,reprint,noshowpacs,superscriptaddress]{revtex4-1}   
\usepackage[T1]{fontenc}	
\usepackage[latin9]{inputenc}	
\usepackage{geometry}		
\usepackage{graphicx}
\usepackage[above,below]{placeins}	
\usepackage{times}
\usepackage{hyperref}  
\hypersetup{colorlinks=true,urlcolor=blue,citecolor=blue,linkcolor=blue}   
\begin{document}

\title{Interview validation of the Physics Lab Inventory of Critical thinking (PLIC)}

\author{Katherine N. Quinn}
\affiliation{Physics Department, Cornell University, 144 East Ave, Ithaca, NY, 14850}
\author{Carl Wieman}
\affiliation{Physics Department, Stanford University, 382 Via Pueblo Mall, Stanford, CA, 94041}
\affiliation{Graduate School of Education, Stanford University, 382 Via Pueblo Mall, Stanford, CA, 94041}
\author{N.G. Holmes}
\affiliation{Physics Department, Cornell University, 144 East Ave, Ithaca, NY, 14850}

\begin{abstract}
Although an important goal of introductory physics labs is to train students in scientific reasoning and critical thinking, currently there are no standard tests in physics designed to assess such skills. We are in the process of developing and validating the Physics Lab Inventory of Critical thinking (PLIC),  an assessment to probe students' critical thinking abilities in physics lab courses. The instrument asks students to critique a set of experimental methods and data and use them to evaluate a particular physical model (the period of a mass on a spring). Currently, we are validating the closed-response survey through interviews with students and present the results of 12 such interviews here. We describe a trend that has emerged from these interviews, with students' reasoning falling into three  main patterns of behavior: selecting all options, cuing on keywords, and critically analyzing. We have found ways to shift students to the last and more desirable behavior. We discuss ways in which these findings are likely relevant to the design of other concept inventories.
\end{abstract}

\maketitle

\section{Introduction}
Opinions on the goals of labs in introductory physics courses vary from reinforcing concepts, to gaining experience with experimental equipment, to developing critical thinking skills~\cite{roles}. Regardless of intended lab goals, it is important to quantitatively measure the effectiveness of different pedagogical strategies. While there are a variety of reserach-based instruments available to test students' conceptual physics understanding, there are none for assessing critical thinking skills for use in introductory physics labs where such skills might be taught. For instance, the Lawson Classroom Test of Scientific Reasoning does not test skills appropriate for college level physics classrooms, not was it statistically validated during its development.~\cite{Lawson} Our instrument, the Physics Lab Inventory of Critical thinking (PLIC) is intended to fill this void.  We define critical thinking to be the ability to ``critique data, identify whether or not conclusions are supported by evidence, and distinguish a significant effect from random noise and variability"~\cite{CriticalThinking}.

The PLIC is a choose-many closed-response survey that asks students to critique the experimental methods and data of two hypothetical groups of students who are testing a model of a mass on a spring. We present the results of 12 interviews as part of our preliminary results, conducted to validate the degree to which the PLIC can evaluate critical thinking skills. We describe three different patterns of thinking behavior observed by students answering the assessment: (1) Selecting all responses without prioritizing answers; (2) cuing into key words learned in class such as human error and percent difference; (3) a discerning pattern of behavior where various answers are carefully weighed and prioritized. We use these results to argue for the validity of the PLIC and guide its development, and more generally to provide insight into the behavior of students taking a multiple-response test. The PLIC is intended to assess how students think critically (not whether or not they do), and so we use the results of these interviews to modify the PLIC in order to prompt the students to be discerning in their answers.

\section{Developing the assessment}
\label{sec:Develop}

\begin{figure}

\begingroup%
  \makeatletter%
  \providecommand\color[2][]{%
    \errmessage{(Inkscape) Color is used for the text in Inkscape, but the package 'color.sty' is not loaded}%
    \renewcommand\color[2][]{}%
  }%
  \providecommand\transparent[1]{%
    \errmessage{(Inkscape) Transparency is used (non-zero) for the text in Inkscape, but the package 'transparent.sty' is not loaded}%
    \renewcommand\transparent[1]{}%
  }%
  \providecommand\rotatebox[2]{#2}%
  \ifx\svgwidth\undefined%
    \setlength{\unitlength}{225.29337706bp}%
    \ifx\svgscale\undefined%
      \relax%
    \else%
      \setlength{\unitlength}{\unitlength * \real{\svgscale}}%
    \fi%
  \else%
    \setlength{\unitlength}{\svgwidth}%
  \fi%
  \global\let\svgwidth\undefined%
  \global\let\svgscale\undefined%
  \makeatother%
  \begin{picture}(1,1.12886433)%
    \put(0,0){\includegraphics[width=\unitlength,page=1]{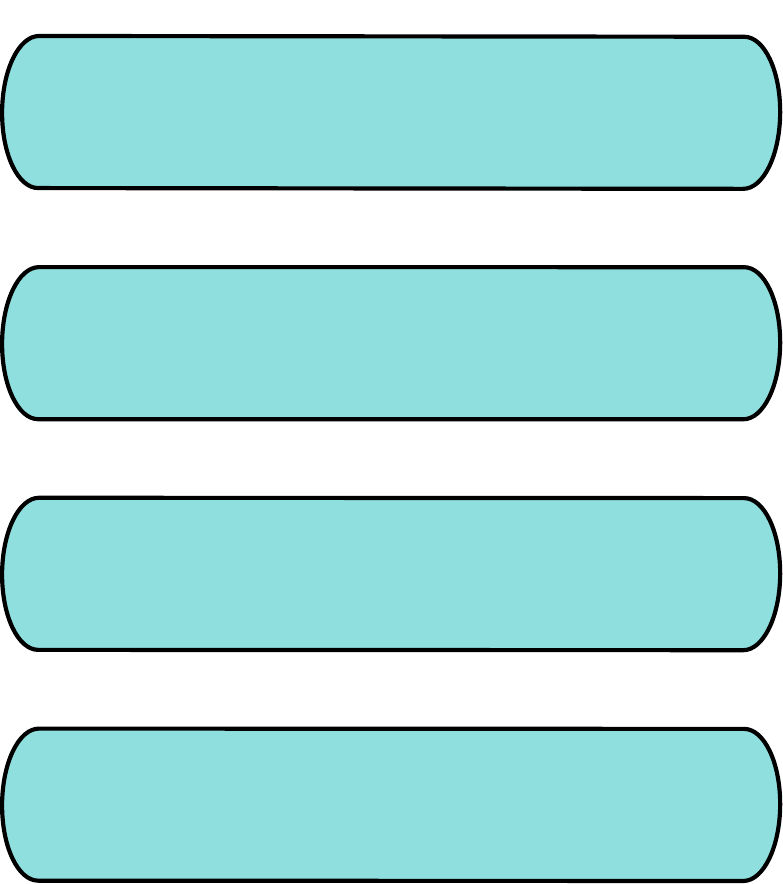}}%
    \put(0.18598241,1.00034891){\color[rgb]{0,0,0}\makebox(0,0)[lb]{\smash{Delineation of the purpose of the test.}}}%
    \put(0.16631568,0.67384174){\color[rgb]{0,0,0}\makebox(0,0)[lb]{\smash{Development of open-response questions.}}}%
    \put(0.14363919,0.94638022){\color[rgb]{0,0,0}\makebox(0,0)[lb]{\smash{Choose topic, create preliminary questions.}}}%
    \put(0,0){\includegraphics[width=\unitlength,page=2]{phases.pdf}}%
    \put(0.35203397,0.71783992){\color[rgb]{0,0,0}\makebox(0,0)[lb]{\smash{Student interviews.}}}%
    \put(0.15473843,0.42674533){\color[rgb]{0,0,0}\makebox(0,0)[lb]{\smash{Development of closed-response questions.}}}%
    \put(0.20345095,0.37927555){\color[rgb]{0,0,0}\makebox(0,0)[lb]{\smash{Field testing and student interviews.}}}%
    \put(0.14050352,0.3344384){\color[rgb]{0,0,0}\makebox(0,0)[lb]{\smash{Consultations with experts, develop scoring.}}}%
    \put(0.15095863,0.12551141){\color[rgb]{0,0,0}\makebox(0,0)[lb]{\smash{Assemble/evaluate test for operational use.}}}%
    \put(0.2319121,0.07804141){\color[rgb]{0,0,0}\makebox(0,0)[lb]{\smash{Validation and reliability testing.}}}%
    \put(0.28098006,0.03320426){\color[rgb]{0,0,0}\makebox(0,0)[lb]{\smash{Consultations with experts.}}}%
    \put(0.28704853,0.62555531){\color[rgb]{0,0,0}\makebox(0,0)[lb]{\smash{Student written responses.}}}%
    \put(0,0){\includegraphics[width=\unitlength,page=3]{phases.pdf}}%
    \put(0.42174918,0.1944688){\color[rgb]{0,0,0}\makebox(0,0)[lb]{\smash{Phase 4}}}%
    \put(0,0){\includegraphics[width=\unitlength,page=4]{phases.pdf}}%
    \put(0.42230185,0.49071725){\color[rgb]{0,0,0}\makebox(0,0)[lb]{\smash{Phase 3}}}%
    \put(0,0){\includegraphics[width=\unitlength,page=5]{phases.pdf}}%
    \put(0.42230185,0.78430273){\color[rgb]{0,0,0}\makebox(0,0)[lb]{\smash{Phase 2}}}%
    \put(0,0){\includegraphics[width=\unitlength,page=6]{phases.pdf}}%
    \put(0.42481593,1.07979045){\color[rgb]{0,0,0}\makebox(0,0)[lb]{\smash{Phase 1}}}%
  \end{picture}%
\endgroup%

\caption{Development framework for the PLIC, following guidelines from~\citep{InstrumentValidation} and validation procedure from~\cite{WilcoxPollockValidation}, with the four main phases of the process and arrows indicating the iterative nature of development where we may return to previous steps depending on findings. Currently, we are in Phase 3, using student interviews to validate the closed-response format.\label{fig:DevProcess}}
\end{figure}

The development framework for this assessment is illustrated in Fig.~\ref{fig:DevProcess}, and was established using protocol from~\citep{physicsAstroAssessment,InstrumentValidation} and further explained for the purpose of the PLIC in~\cite{initialPaper}. Because critical thinking of this sort is context dependent~\cite{criticalThinkingTeaching}, assessing these skills requires a context-specific tool. Questions in the PLIC ask students to interpret (fictitious) data, critique experimental design, and evaluate model breakdown when conflicting data are presented. Specifically, the model in question is a simple mass on a spring, chosen because the physics content and experimental set-up should be familiar to any student in an introductory physics course.

The initial data were created by a physicist conducting the actual experiment. The basis for the question sequence followed the physicist's self-questions related to their choice of methods, and data collection and interpretation. To develop and refine the free-response version, interviews with students were conducted, resulting in major changes such as modification of data and the addition of questions and sections~\citep{initialPaper}. The free-response survey was given to students in different courses and their answers were categorized to convert the survey to a closed format. Several iterations of the closed-response PLIC have emerged, primarily as a result of interviews with students and consultations with expert physicists. Answers to questions are intended to capture all probable responses given by students. 

\subsection{Current assessment format}
\label{subsec:Current}

The current structure of the PLIC is as follows: Two fictitious groups of physicists design and carry out an experiment to explore the Hooke's law model of simple harmonic motion, where the period of oscillation of a mass on a spring is given by $T = 2\pi\sqrt{\frac{m}{k}}$. Group~1 examines two masses and numerically compares the resulting spring constants, whereas Group~2 examines multiple masses and plots $T^2$ (period of oscillation squared) versus $m$ (the mass) to graphically compare the results. According to the model, this should be a linear relationship with the best fit line passing through the origin, however collected data appear to contradict this. The instrument proposes that an intercept offset be added to the graph and then questions the student about this new fit. Through closed-response questions, students are asked to assess the data, recommend next steps, and compare the methods and results of the two groups.

\begin{figure}
\fbox{\includegraphics[width=0.9\linewidth]{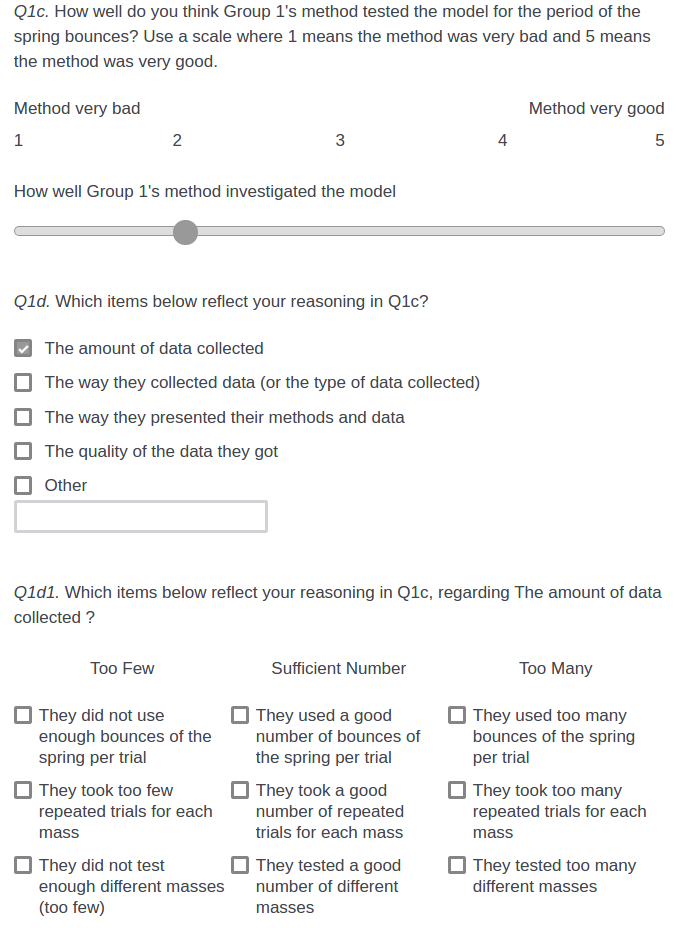}}
\caption{Sample questions from the PLIC. Students are asked to assess a particular aspect of the fictituous experiment regarding a mass on a spring, in this case the method used, and then provide reasoning for their assessment. As there are many possible reasons, answers are catagorized and, depending on student selection, new questions appear. Here, the bottom question appeared after the answer ``amount of data collected" was selected. This allows for multiple lines of thought and reasoning. Ideas, such as the number of trials per mass, are organized by row with different columns grouping ranges of each idea for a more organized presentation. \label{fig:SampleQ}}
\end{figure}

Example questions from the survey are presented in Fig.~\ref{fig:SampleQ}. The instrument assesses students' reasoning and critical thinking by asking them to make a judgment and justify their answer. A traditional format of ``choose one multiple choice" would be inadequate because there is no single right answer for critical thinking, as it is multi-layered and depends on logical evaluations. The design and format were modelled after Wilcox and Pollock's coupled multiple-response assessment~\cite{WilcoxPollockCoupled}. For example, students are asked to assess how well a particular method tests the model and the following question asks them to provide reasons for their selection. Once they make a selection such as ``the amount of data collected", additional questions appear which probe their reasoning in more detail, as illustrated in Fig.~\ref{fig:SampleQ}. This dynamic flexibility in format is possible through the online survey tool used for our purposes, Qualtrics.

\section{Methods}
\label{sec:Methodology}

Twelve think-aloud interviews were conducted individually with one interviewer, and audio-recorded. Students were asked to complete the closed-response survey while vocalizing their thoughts and explaining their choices, with the interviewer interjecting as needed, and after which structured follow-up questions were asked. It took between 18 and 45 minutes for the students to complete the survey. The time varied mostly by individual and familiarity with the subject material. For instance, those unfamiliar with residual plots took additional time to read the descriptions and familiarize themselves, while others were able to proceed directly to the questions. Students were financially compensated for their time.

Detailed notes on each interview were compiled and examined to determine underlying themes and common behaviors. Interviews highlighted sections of the survey where concepts were insufficiently described (such as residual plots), and as a result the small modifications to the wording of the PLIC were made after the majority of interviews. A breakdown of demographic characteristics of interview sample is presented in Table~\ref{tab1}.

\begin{table}[htbp]
\caption{Demographic breakdown of all 12 students interivewed.\label{tab1}}
\begin{ruledtabular}
\begin{tabular}{c l l }
 \textbf{Category} & \textbf{Breakdown} & \textbf{N}\\ 
 \hline
Major & Physics/Applied and engineering physics & 8 \\
      & Non-declared STEM & 2 \\
      & Other declared STEM & 2 \\
\hline
Academic Level & Freshman & 6 \\ 
               & Sophomore & 3 \\
               & Junior & 1 \\ 
               & Graduate Student & 2 \\
\hline
Gender & Female & 6 \\ 
       & Male & 6 \\
\hline
Race/Ethnicity & African-American & 2 \\
               & Asian/Asian-American & 3 \\ 
               & White & 5 \\ 
               & Other & 2
\end{tabular}
\end{ruledtabular}
\end{table}

\section{Interview Results and patterns of responses}
\label{sec:Interview}

The goal of the PLIC is to measure critical thinking skills, not general physics knowledge, and so great care was taken to minimize the need for prior physics concept knowledge. Through interviews conducted with non-physics majors who have taken at least one physics course, we were able to conclude that although some exposure to physics is necessary, the need for prior physics content knowledge is minimal and appropriate for assessing learning in the context of a physics lab course.

A deliberate choice was made to not include a ``No Opinion" response. Although it is conceivable that they may have no opinion, it has been shown that a ``no opinion" option can reduce the quality of responses to closed-response surveys, since it discourages students from performing the cognitive work necessary to answer~\citep{noOpinion}. At no point during the validation process did a student indicate a preference for such an option, and so it was decided that the benefit of forcing students to select an answer outweighed the benefits of having this additional possible response.

Through these interviews we identified different patterns in thinking and reasoning during the test, which we classified into three main behaviors. We discuss these behaviors and their implications below.

\subsection{Thinking behavior: Selecting all options}

Three students, when presented with the multitude of possible answers, proceeded to select all options that resonated with them regardless of priority or importance. For instance, questions regarding what Group 1 should do next prompted one student to remark ``these are all good ideas" and to select almost all options.

Two such students had preferences for options related to human error and increasing the amount of data collected (e.g. number of trials per mass, and number of masses considered), and use of better equipment. For questions related to ``next steps" for each group, these students selected almost all options, except for ``wrap up their experiment". The final section of the assessment asks students to compare Group 1 and Group 2 methodology and data. Both students who selected all options had a preference for few measurements with increased precision (Group 1) over a greater number of less precise measurements (Group 2). Deviation from the model was attributed to human error or random fluctuation, and one student ignored the model breakdown.

Compared with a conceptual diagnostic test, closed-response options in the PLIC are all in some way ``correct". This behavior therefore does not represent critical thinking, as the students are not being critical of the many possible responses. For this reason, the PLIC has been modified to allow no more than three answers per question. When urged by the interviewer to select no more than three options after this behavior was identified during the interview, students were forced to be critical of their choices and prioritizing their answers. This needed to be repeated several times, leading to a modification of the PLIC which notifies students when they select too many options.

\subsection{Thinking behavior: Cuing}

Five students primarily selected options with keywords learned from class, such as the percent error between two measurements and ``human error". This behavior was most common when answering the set of questions related to Group 1 methodology (as described in Section~\ref{subsec:Current}). When the question used vocabulary and terminology familiar to students, several defaulted to this particular pattern.

One student who exhibited this behavior noted that the initial questions are ``just plug and chug stuff" and proceeded to select only answers containing key words from class. Later in the interview, as the questions became more unfamiliar, the same student noted that they ``didn't think about [doing] that", and their thinking behavior changed to be more discerning. All students who were unfamiliar with the material later in the interview exhibited the switch from cuing to discerning behavior, further discussed in Section~\ref{sec:Discussion}.

\subsection{Thinking behavior: Discerning} 

Ten of the interviewees, at some point, displayed evidence of critically evaluating the question and the possible answers. This critical thinking manifested as considering all options presented to them yet selecting only a small subset of those options. The sophistication of the critical thinking differed between students and was related to the options they selected. Some focused on elements that would reduce human error, while
others noted the breakdown of the model and suggested this be further explored. 
Five discerning students would change their answer to the initial judgment question after justifying their answer. They argued that the responses contained reasons they had not initially considered. This demonstrates that seeing the answer options affects students' responses, but we argue that this triggers students to think critically. Given that there was variability in these discerning students' responses, we argue that this feature (prompting students to think critically) allows the PLIC to assess the quality of students critical thinking, as intended.

Of particular note is that students would switch to discerning behavior when the material was presented in an unfamiliar way (such as with Group 2 data and methods) or if they were told by the interviewer to select no more than three answers to a particular question.

\section{Discussion and future direction}
\label{sec:Discussion}

In this paper we present the results of 12 interviews conducted to validate the closed-response version of the PLIC. We noticed three different thinking patterns: (1) Selecting all responses, regardless of importance or priority, (2) cuing behavior, where students cue into keywords, and (3) discerning behavior, where students prioritize and select relevant answers. It is very likely that these behavior patterns are common to assessments which allow multiple responses, and so their relevance extends beyond the PLIC. In particular, we use these results to guide the PLIC structure to prompt the students to be discerning in their answers, in order to assess $how$ (not whether) they think critically.

Critical thinking was exhibited by the students primarily when they were discerning in their choices. This behavior can be prompted in several ways, which have been used to guide the format and structure of the PLIC. First, by excluding ``no opinion" responses, students needed to perform the cognitive work necessary to come up with an answer~\cite{noOpinion}. Second, by having the number of possible answers limited to less than three, students who were selecting all options were forced to prioritize their answers. Finally, by presenting familiar concepts in new or unfamiliar ways, such as graphing data with residual plots, students who were initially cuing to key words learned from class switched to carefully weighing the available options. In particular, this occurred when students were answering questions regarding the fictitious data and experimental methods of Group 2, indicating that this portion of the PLIC is more useful for evaluating critical thinking. Once students exhibit this discerning behavior, their answers do differ with regards to which topics they prioritize (minimizing human error, or investigating the model breakdown as data contradict predictions). These different answers, triggered when students are thinking critically, can be used to distinguish between more novice-like and expert-like thinking within the context on labs and will be a focus in the next phase of PLIC development. From this, we can claim that the structure of the PLIC allows us to effectively probe the quality of students' critical thinking. Furthermore, these techniques could be valuable in the development of other multiple-response assessments.

The PLIC will be compared with expert responses to better quantify differences between expert-like and novice-like reasoning. This will provide the basis for a scoring scheme, where answers most similar to expert-like score higher than those more dissimilar. Once this is complete, we will enter the final development phase (as show in Fig.~\ref{fig:DevProcess}) and assemble a version of PLIC for large scale, operational use.

\acknowledgments{This material is based upon work supported by the National Science Foundation under Grant No. DUE-1611482-01. We would also like to thank Prof. Bonn of UBC for providing the initial data that formed the basis of the PLIC and for his help and input during the development, and Emily Smith for valuable input.}

\bibliography{PERC2017_paper}

\end{document}